\documentclass[prl,twocolumn,amsmath,secnumarabic,amssymb]{revtex4-1}

\usepackage{graphicx}
\usepackage{color}
\begin{document}


\title{Experimental demonstration of a dusty plasma ratchet rectification and its reversal}


\author{Ya-feng He$^{1}$}
\email[Email:]{heyf@hbu.edu.cn}
\author{Bao-quan Ai$^{2}$}
\email[Email:]{aibq@scnu.edu.cn}
\author{Chao-xing Dai$^{1}$}
\author{Chao Song$^{1}$}
\author{Rui-qi Wang$^{1}$}
\author{Wen-tao Sun$^{1}$}
\author{Fu-cheng Liu$^{1}$}
\author{Yan Feng$^{3}$}
\email[Email:]{fengyan@suda.edu.cn}

\affiliation{$^1$Hebei Key Laboratory of Optic-electronic Information Materials, College of Physics Science and Technology, Hebei University, Baoding 071002, China\\
$^2$Guangdong Key Laboratory of Quantum Engineering and Quantum Materials, SPTE, South China Normal University, Guangzhou 510006, China\\
$^3$Center for Soft Condensed Matter Physics and Interdisciplinary Research, School of Physical Science and Technology, Soochow University, Suzhou 215006, China.}


\begin{abstract}

The naturally persistent flow of hundreds of dust particles is experimentally achieved in a dusty plasma system with the asymmetric sawteeth of gears on the electrode. It is also demonstrated that the direction of the dust particle flow can be controlled by changing the plasma conditions of the gas pressure or the plasma power. Numerical simulations of dust particles with the ion drag inside the asymmetric sawteeth verify the experimental observations of the flow rectification of dust particles. Both experiments and simulations suggest that the asymmetric potential and the collective effect are the two keys in this dusty plasma ratchet. With the nonequilibrium ion drag, the dust flow along the asymmetric orientation of this electric potential of the ratchet can be reversed by changing the balance height of dust particles using different plasma conditions.

\end{abstract}



\maketitle
\indent A dusty plasma (or complex plasma) consists of micron-sized dust particles immersed in a plasma \cite{Morfill, Shukla1}. These dust particles are highly charged so that they can be strongly coupled. These unique properties make the dusty plasma an ideal model system for phase transitions, transport processes, and nonlinear waves \cite{Chu,Rubin,Mamun,Kalman,Thomas, Wong, Thomas2, Hyde, Resendes, Hartmann,Feng2, Ott, Wang, Killer, Joyce, Thomas, Melzer0, Ivlev}.

\indent Feynman ratchets \cite{Feynman, hanggi}, composed of periodically arranged units with asymmetry, have been experimentally demonstrated in a few physical \cite{Skaug, Roeling, Silva}, chemical \cite{Wilson}, and biological \cite{Park} systems. The Feynman ratchet model provides a fundamental physics concept rectifying the nonequilibrium fluctuations into directional particle motion. In particular, the direction of particle flow is reversible as changing the control parameters. However, this reversibility remains an open issue.

\indent Here, we extend the idea of the Feynman ratchet to the dusty plasma system to experimentally demonstrate a dusty plasma ratchet rectification and its reversal. The dust particles are rectified into a directional flow with desirable direction by regulating the gas pressure or the power of the radio-frequency (rf) plasma. Using numerical simulations, we further explore the underlying mechanisms of the rectification and its reversal. We find that the particle flow along an electric potential of the ratchet is induced by the local ion drag inside the sawteeth until it is balanced by the neutral drag, and the direction of particle flow follows the asymmetric orientation of the electric potential of the ratchet that can be controlled, or even reversed, by changing the balance height of dust particles with different plasma conditions.

\begin{figure}[htbp]
  \begin{center}
  \includegraphics[width=7cm,height=5cm]{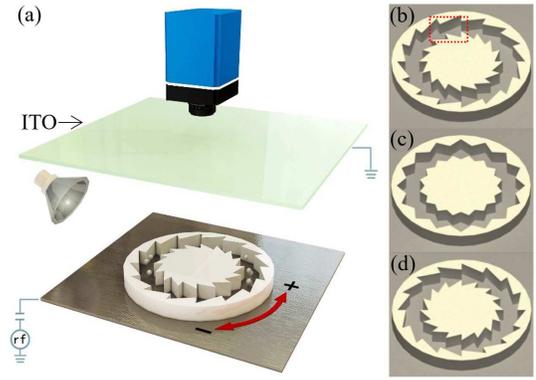}
  \caption{(color online). Sketch of the dusty plasma ratchet experiment (a). Dust particles can perform a specific directional motion along the saw channel enclosed by the inner and outer gears. Here, we define the positive and negative motions as `+' and `-' shown. (b)-(d) show gears with different orientation on the sawtooth. The asymmetric orientation on the sawtooth in (b) is opposite to that in (d). The sawtooth in (c) is symmetric along the saw channel. The field of view of our data analysis is marked by using the dotted rectangle in (b). Dust particles are suspended $\sim5-9$ mm above the lower electrode under different conditions. ITO: Indium Tin Oxides glass plate.}
  \end{center}
\end{figure}

\indent A circular resin gear (named inner gear) with asymmetric sawtooth is placed concentrically with another circular resin gear (named outer gear) on a horizontal lower electrode, as shown in Fig.~1(a). The orientation of the sawtooth in the inner gear is the same as that in the outer gear. As a result, the inner and outer gears enclose a saw channel, and the width of the saw channel changes periodically. The dimensions of the gears are described in detail in~\cite{simulations}.

\indent Argon plasma is produced in a vacuum chamber by the capacitive coupling the lower electrode to a rf power source ($13.56$ MHz) via a capacitor. Then, we introduce macroporous cross-linked polystyrene microspheres (dust particles) into one side of the saw channel by using a capillary glass tube. The radius of the dust particle is $r_d = 11.5$ $\mu$m and the mass density is $\sim0.7$ g/cm$^3$ as reported by the manufacturer. They are illuminated by a $18$ W flood lamp from the side and imaged by a camera from the upper transparent electrode. After the motion of dust particles is recorded, we use the moment method \cite{Feng} to calculate positions of dust particles.

\begin{figure}[htbp]
  \begin{center}\includegraphics[width=8cm,height=9.2cm]{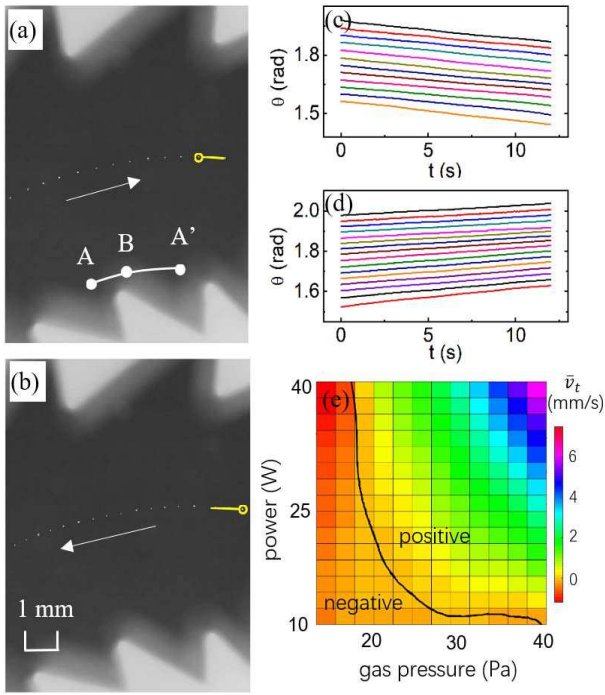}
  \caption{(color online). Snapshots of dust particles in the negative (a) and positive (b) flows of dust particles, indicated by arrows, in the same gear structure, for the experiment conditions of (35 Pa, 10 W, peak-to-peak voltage $V_{pp}=41$ V) and (40 Pa, 10 W, $V_{pp}=39$ V), respectively. In (a) and (b), a typical trajectory of one selected dust particle during 12 seconds is shown using the solid curve with a starting point marked as a circle. The labels ``A'', ``B'', and ``A' '' in (a) mark the positions of the maximum, minimum, and maximum of electric potential in the arc (radius $r$ and length $l= r \Theta $), respectively. The angular positions of dust particles for the negative (c) and positive (d) flows of dust particles. The mean tangential velocity $\bar{v}_t$ of the directional motion (e) as functions of the gas pressure and the rf power. The critical curve of $\bar{v}_t = 0$ is indicated by the solid curve. The total number of dust particles in this experiment is $N = 204$.}
  \end{center}
\end{figure}

\indent As our main result of this Letter, we observe the steady directional motion of dust particles along the saw channel in our experiment, as shown in Fig. 2 and the Supplemental Material \cite{Supplementary}. When these dust particles are introduced into the plasma, due to their Coulomb repulsion, they arrange themselves into a circular dust chain along the center of the saw channel first. After about several seconds, the dust chain rotates persistently along the saw channel, in the direction indicated in Fig.~2. Our experiment clearly demonstrates that our dusty plasma ratchet realizes the rectification of particles in plasmas.

\indent Remarkably, we discover that the direction of the steady motion can be modified by changing the experimental conditions as shown in Figs.~2(a)-2(d). We experimentally realize both negative and positive flows of dust particles by changing the plasma conditions only, as shown in Fig.~2(e). These experimental results clearly demonstrate that, in dusty plasmas, one can achieve bidirectional rectification without changing the symmetry of the sawtooth. Note that, in our experiment, the rotation speed of the dust chain can be modified by the plasma conditions, the gear parameters, and the number of dust particles, as we present one by one next.

\indent Figure 2(e) illustrates the behavior of our dusty plasma ratchet with 204 dust particles, over a wide range of plasma conditions. The directional motion of the dust chain is described using the mean tangential velocities $\bar{v}_t$. It can be seen that, at larger gas pressure and higher rf power, the mean tangential velocity of the directional motion can be as high as $\approx$ 7 mm/s, with the corresponding rotation period of the dust chain is $\approx$ 14 s, for the plasma conditions of 40 Pa and 40 W. In Fig.~2(e), we draw a solid curve indicating the separation of the negative and positive flows of dust particles. When we set the gas pressure and rf power around the critical curve, the dust particles stop the directional motion and only oscillate around their equilibrium positions in the saw channel. From our observation, the transition between the negative and positive flows of dust particles is reversible by changing the gas pressure and rf power.

\indent Next we experimentally verify two keys to the rectification of dust particles, the asymmetric potential and the collective effect.

\indent The asymmetry of the sawtooth plays one key role in the dust particle rectification. Around the gears of our experiments, the distribution of electric potential exhibits asymmetry along the saw channel (tangential direction). To illustrate the asymmetric electric potential in plasmas, a segment of arc with the radius $r$ and length $l=r\Theta$ ($\Theta=2\pi/n$, $n=24$ is the number of sawtooth of the gear) is drawn in Fig.~2(a). Obviously, the electric potentials at points ``A'' and ``A' '' are higher than that at point ``B'', since ``B'' is  closer to the sheath of the inner gear. Moreover, the distribution of the electric potential along the arc is not symmetric because the length $l_{AB}$ is smaller than $l_{BA'}$. This leads to an asymmetric potential, as we verify in~\cite{simulations}.

\indent To further confirm the key role of the electric potential of the ratchet in the rectification of dust particles, we flip the gears up-down as in Fig.~1(d), and we find that, under the same gas pressure and rf power, the direction of particle flow reverses, and the rotation speed is almost unchanged for a similar number of dust particles inside the saw channel. We also verify that symmetrical gears, Fig.~1(c), would not cause any net flow of dust particles, for any conditions we tried.

\begin{figure}[htbp]
  \begin{center}\includegraphics[width=7cm,height=4cm]{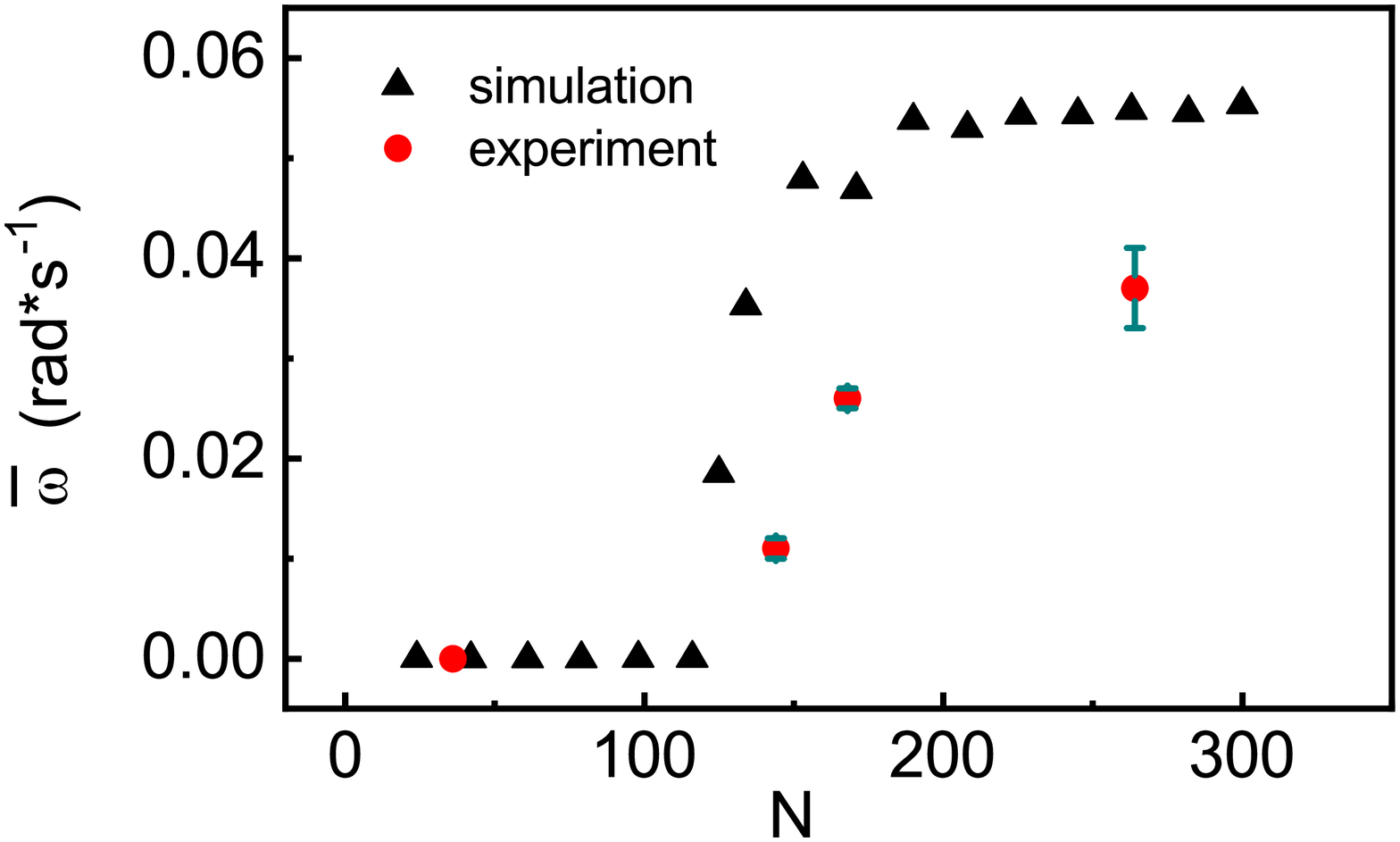}
  \caption{The dependence of the mean angular speed on the total number of dust particles. In experiments, multilayers of dust particles form in the saw channel when $N$ more than about 300. The plasma conditions are 45 Pa, 11 W, 45~V $V_{pp}$.}
  \end{center}
\end{figure}

 \indent The collective effect of the interparticle interaction plays the other key role in the rectification of these charged dust particles. When the number of dust particles is small, a few dust particles are confined within the potential well of one sawtooth, so that we do not observe directional flow of dust particles, i.e., the mean angular speed $\bar{\omega}$ $=$ $\sum_{i=1}^{N}\omega_{i}/{N}$ $=$ $0$ rad/s, as in Fig.~3. However, when the dust particles are more and more, the region of the potential well of one sawtooth is not big enough to contain these charged dust particles, so that they move outside the potential well of this sawtooth to form a single circular chain with other dust particles from the potential wells of other sawteeth. At this moment, the collective effect works, and the directional flow of dust particles occurs. The mean angular speed increases with the number of dust particles as in Fig.~3.

\indent However, in our experiments, more than about 300 introduced dust particles will cause multilayers of dust particles inside the saw channel. As a result, the ion wake effect in the plasma sheath~\cite{Melzer0, Piel} would substantially modify the dynamics of dust particles. Here, we only focus on the dynamics of a single layer of dust particles.

\indent The collective effect of the particle repulsion has been found in the organic electronic ratchet \cite{Roeling} and the superconducting ratchet \cite{Silva}. In our experiments, the dust particles are strongly coupled due to their high charges $Q$ $\sim$$-9.5\times10^{4}$ $e$ (a typical value from~\cite{simulations}, where $e$ is the elementary charge), which is close to that calculated from empirical formula $Q$ $=$ $-1400 r_{d}T_{e}$ $e$ $\sim$ $-6.9\times10^{4}$ $e$ \cite{Bonitz1}. The high charges of dust particles are capable of enhancing this collective effect to speed up the flow of dust particles.

\begin{figure}[htbp]
  \begin{center}\includegraphics[width=9.3cm,height=8.3cm]{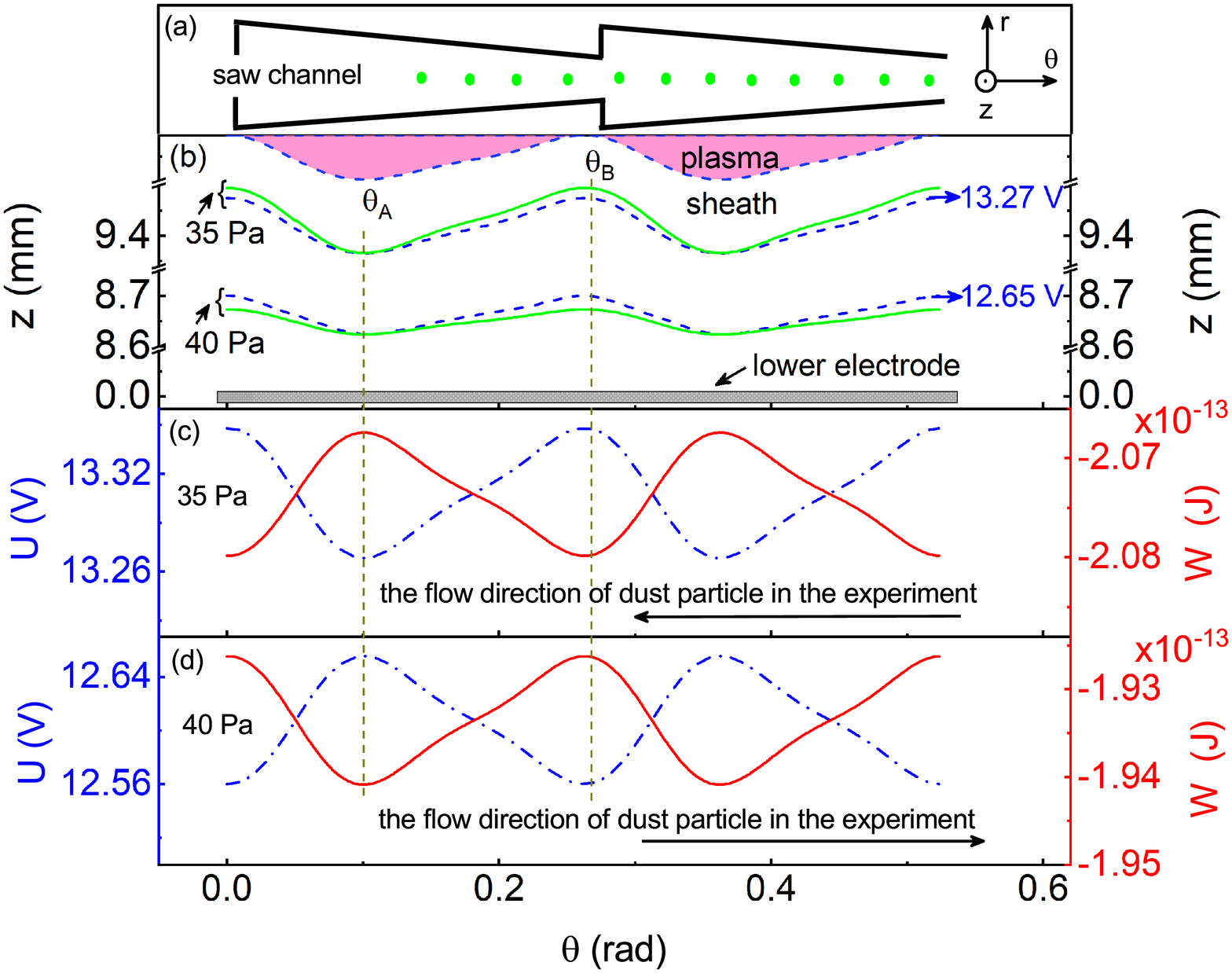}
  \caption{(color online). Sketch of the saw channel (a) with two sawteeth, with points indicating dust particles. (b): Balance heights of dust particles above the lower electrode (solid curves in which $mg=-Q\frac{\partial U}{\partial z}$) and the dashed equipotential curves of $U$=12.65 and 13.27 V in the sheath, at 35 and 40 Pa, respectively, from our COMSOL simulations~\cite{simulations}. The solid curves of the balance height of dust particles don't overlap with the dashed equipotential curves, indicating that the electric potential at the balance height of dust particles is varied. Electric potential $U$ (dash-dotted curve) and electric potential energy $W=QU$ (solid curve) at the balance height of dust particles under the conditions of 35 Pa (c) and 40 Pa (d). The asymmetric orientation of the electric potential energy for 35 Pa is opposite to that for 40 Pa, which reasonably results in the reversal of particle flow. Note that the plasma rf power is 10W, and $V_{pp}=41$ V, $n_{e,i}$$\sim$ $1.1\times10^{15}$ m$^{-3}$, $T_e$$\sim$ 4.4 eV for (c), while $V_{pp}=39$ V, $n_{e,i}$$\sim$ $1.5\times10^{15}$ m$^{-3}$, $T_e$$\sim$ 4.3 eV for (d).}
  \end{center}
\end{figure}

\indent We perform numerical simulations to verify our experimental findings. First, the plasma parameters such as the electric potential $U$ inside the whole saw channel is obtained using COMSOL simulations \cite{simulations, COMSOL}. Next, we consider the confinement of the charged dust particles in the $r$, $z$, and $\theta$ directions, as in Fig.~4(a), respectively. In the radial $r$ direction, dust particles are confined to the center of the saw channel by the gears, as in Fig.~4(a), which agrees with the experimental trajectories as in Figs.~2(a), 2(b). Then, we just focus on the motion of dust particles in the $z-\theta$ space along the center of the saw channel. In the vertical $z$ direction, the thickness of the sheath changes periodically with the width of the saw channel \cite{Kim}, as indicated by the dashed equipotential curves in Fig.~4(b). The dust particles are suspended above the lower electrode at the balance height, where $mg=-Q\frac{\partial U}{\partial z}$. From Fig. 4(b), the balance height of dust particles varies periodically along the saw channel.

\indent We find that the variation of the balance height of dust particles does not overlap with the equipotential curves in the sawteeth at 35 Pa and 40 Pa as in Fig.~4(b). This is reasonable because the balance height of dust particles is determined by the vertical potential gradient $\frac{\partial U}{\partial z}$, not $U$. Thus, the electric potential at the balance height of dust particles also varies periodically. Figures~4(c) and 4(d) show the electric potential $U$ and the energy $W=QU$ at the balance heights of dust particles in Fig.~4(b) for two conditions of 35 Pa and 40 Pa, respectively. They both exhibit an asymmetric potential within each sawtooth along the saw channel.

\indent The conservative electric potential of the ratchet itself cannot lead to a directional motion of dust particles. Without other external forces, a single dust particle will definitely go to the potential minimum from any locations, as in Fig.~5(b). However, the electric potential of the ratchet along the saw channel can induce an azimuthal ion drag force $F_{i\theta}$ which can drive dust particles through the ion-dust collision. Due to this $F_{i\theta}$, a single dust particle can be pushed away from the potential minimum to a new location on the slope of the potential well of one sawtooth, as indicated in Fig.~5(b).

\begin{figure}[htbp]
  \begin{center}\includegraphics[width=9cm,height=5.5cm]{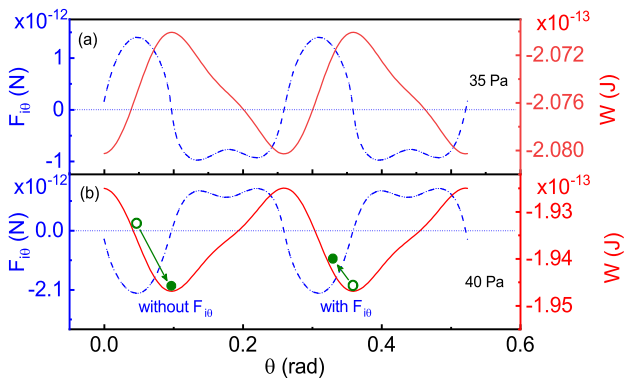}
  \caption{(color online). Azimuthal ion drag force $F_{i\theta}$ of one dust particle (dashed curve) and electric potential energy $W=QU$ (solid curve) at the balance height of dust particles. For the condition of 35 Pa in (a), $f_{i\theta}<0$; while for 40 Pa in (b), $f_{i\theta}>0$. In (b), it is also illustrated how a single dust particle moves from a certain initial position (hollow circle) to steady position (filled circle) without/with the ion drag force $F_{i\theta}$ in the potential well.}
  \end{center}
\end{figure}

\indent When the number of dust particles is large enough to fill up the potential well of one sawtooth until a dust chain forms, a dust particle could climb more easily over the potential barrier of one sawtooth along the slanted side rather than the steep side, with the help from other particles via the interparticle repulsion, resulting in a persistent dust flow. In the potential well of each sawtooth, although the direction of $F_{i\theta}$ on the slanted side of the potential well  is opposite to that on the steep side as shown in Fig.~5, for the whole circular dust chain, a net ion drag force could appear due to the asymmetry of the electric potential of the ratchet. To characterize the collective effect of the whole dust chain, we define a net ion drag force as $f_{i\theta}$$=$$\frac{\oint_{l} \textbf{F}_{i\theta} \cdot d \textbf{l}}{l}$, i.e., a circulation of $F_{i\theta}$ divided by the circular path of integration $l$ following the dust chain. This nonequilibrium driving force of net ion drag could be non-zero, in the order of $\sim$$10^{-13}$ N, which can overcome the drag force from the neutral gas, also about $\sim$$10^{-13}$ N, to generate a steady dust flow.

 \indent The reversal of particle flow caused by different pressures and powers can be explained by the different variation of the electric potential distribution and the balance height of dust particles. As the gas pressure/power changes, the vertical balance height of dust particles and the electric potential distribution inside the $z-\theta$ space of the sheath are both modified as shown in Fig.~4(b). The variation trends for both solid curves and dashed curves are similar, however, the variation quantities are different. As clearly shown in Figs.~4(b)-4(d), the electric potential at $\theta_{A}$ for 35 Pa is lower than that at $\theta_{B}$, while the electric potential at $\theta_{A}$ is larger than that at $\theta_{B}$ for 40 Pa. Therefore, the asymmetric orientation of the electric potential of the ratchet at the balance height of dust particles reverses at these two pressures, as dash-dotted curves shown in Fig.~4(c) and 4(d). As a result, the direction of the net ion drag force is reversed, $f_{i\theta}<0$ ($f_{i\theta}>0$), resulting in the negative (positive) flow of dust chain, and the direction of dust flow is reversed.

\indent Based on the above analyses, we also perform one-dimensional (circular) numerical simulations of dust particles along the saw channel, with the four forces acting on dust particles of the interparticle Yukawa interaction, the electric force, the neutral gas damping, and the ion drag, as described in detail in \cite{simulations}. We obtain both negative (35 Pa, 10 W) and positive (40 Pa, 10 W) flows of dust particles by combining the asymmetric electric potential of the ratchet with the collective effect. We also observe the variation of the rotation speed of the whole dust chain when the number of dust particles is not too small, as shown in Fig.~3. Our numerical results are well consistent with our experimental observations.

\indent In summary, we demonstrate the rectification of dust particles in a dusty plasma ratchet, and the manipulation of the direction of particle flow by changing the gas pressure or the power of the plasma. The asymmetric electric potential and the collective effect play the key roles in the rectification of this dusty plasma ratchet. All experimental observations are verified in our numerical simulations. We show that the particle flow along an electric potential of the ratchet is induced by the local ion drag inside the sawteeth, and the flow direction follows the asymmetric orientation of the electric potential of the ratchet that can be controlled, or even reversed, by changing the balance height of dust particles with different plasma conditions. Our findings could inspire the future studies of controlling particle or even macroscopic object transport, and reversibility in other plasma systems using the similar concept. Our results also suggest that, in one experiment, dust particles with different sizes suspended at different heights could be rectified in opposite directions, so that particle separation might be achieved.

\indent This work is supported by the National Natural Science Foundation of China (Grants No. 11975089, No. 11575064), the Program for National Defense Science and Technology Innovation Special Zone, the Program for Young Top-Notch Talents of Hebei Province, and the GDUPS (2016). Work in Suzhou is supported by the National Natural Science Foundation of China under Grant No. 11875199 and the 1000 Youth Talents Plan. The authors thank two anonymous referees for their constructive suggestions about the physics mechanisms.

\end{document}